\documentclass{article}

\PassOptionsToPackage{numbers, compress}{natbib}
\usepackage[preprint]{neurips} 
\usepackage[none]{hyphenat}
\usepackage[utf8]{inputenc} 
\usepackage[T1]{fontenc}    
\usepackage{hyperref}       
\usepackage{url}            
\usepackage{booktabs}       
\usepackage{amsfonts}       
\usepackage{nicefrac}       
\usepackage{microtype}      
\usepackage[table, dvipsnames]{xcolor} 
\usepackage{graphicx}
\usepackage{array}
\usepackage{wrapfig}
\usepackage[most]{tcolorbox}
\usepackage{soul} 
\usepackage{enumitem}
\usepackage{xspace}
\usepackage{tabularx}
\usepackage{ragged2e} 
\usepackage{amsthm}

\newtheorem{definition}{Definition}
\usepackage{tikz}
\usetikzlibrary{arrows.meta, shapes.geometric, positioning, fit, backgrounds, shadows, mindmap, patterns, shapes, calc}

 \hypersetup{
     colorlinks = true,
     linkcolor = blue,
     anchorcolor = blue,
     citecolor = blue,
     filecolor = blue,
     urlcolor = blue
 }

\title{AI Safety vs. AI Security: \\Demystifying the Distinction and Boundaries}

\author{%
  \textbf{Zhiqiang Lin}\;\;\;
  \textbf{Huan Sun}\;\;\;
  \textbf{Ness Shroff}\\
The Ohio State University \\
  \;\texttt{\{lin.3021,\;\;sun.397,\;\;shroff.11\}@osu.edu}\\
}

\begin{document}
\bibliographystyle{plainnat} 

\maketitle

\begin{abstract}
Artificial Intelligence (AI) is rapidly being integrated into critical systems across various domains, from healthcare to autonomous vehicles. While its integration brings immense benefits, it also introduces significant risks, including those arising from AI misuse. Within the discourse on managing these risks, the terms ``AI Safety'' and ``AI Security'' are often used, sometimes interchangeably, resulting in conceptual confusion. This paper aims to demystify the distinction and delineate the precise research boundaries between AI Safety and AI Security. We provide rigorous definitions, outline their respective research focuses, and explore their interdependency, including how security breaches can precipitate safety failures and vice versa. Using clear analogies from message transmission and building construction, we illustrate these distinctions. Clarifying these boundaries is crucial for guiding precise research directions, fostering effective cross-disciplinary collaboration, enhancing policy effectiveness, and ultimately, promoting the deployment of trustworthy AI systems.
\end{abstract}

\section{Introduction}

Artificial Intelligence (AI) is increasingly embedded in critical societal domains, from autonomous vehicles navigating urban environments and medical algorithms supporting diagnostic decisions, to financial systems executing high-stakes trades~\cite{russell2019human, Marcus2019RebootingAI}. While this integration offers substantial gains in efficiency and capability, it also heightens a range of risks including malicious use by adversaries, unforeseen failures with harmful consequences, and systemic disruptions that are difficult to anticipate or manage~\cite{Hendrycks2023OverviewCatastrophicAIRisks, BengioEtAl2025InternationalAISafetyReport}.


Complicating efforts to address these diverse risks is the inconsistent use of key terminology. In particular, the terms \textit{AI Safety} and \textit{AI Security}, though central to discussions of responsible AI development, are frequently used interchangeably or inconsistently across technical and policy communities~\cite{QiEtAl2024AIRiskFramework}. For example, the recent \textit{International AI Safety Report} adopts a broad interpretation of AI Safety that encompasses risks from malicious use, defining safety to include preventing AI from ``being used for nefarious purposes''~\cite{BengioEtAl2025InternationalAISafetyReport}. This expanded usage reflects both the historical evolution of safety concerns, originally focused on long-term alignment and control of possible 
Artificial General Intelligence (AGI) systems~\cite{Bostrom2014Superintelligence}, and the distinct developmental trajectories of the AI safety and cybersecurity research communities~\cite{QiEtAl2024AIRiskFramework}.

Addressing AI risks effectively, however, requires a clear and consistent understanding of the boundary between these two domains. \textit{AI Safety} and \textit{AI Security} both aim to reduce harm and ensure reliable system behavior, but they originate from different conceptual foundations, target fundamentally different classes of threats, namely \textit{unintentional system failures} versus \textit{intentional adversarial actions}, and rely on distinct methodological ``toolboxes''~\cite{Amodei2016ConcreteProblems}. However, expansive or ambiguous usage of these terms can obscure these critical distinctions, making it difficult to align appropriate technical strategies with the nature of the underlying risk.

The difference between Safety and Security 
can be intuitively illustrated using a classic communication model from computer networks and cryptography:
\begin{itemize}[noitemsep,topsep=0pt]
    \item \textbf{Safety Concern (Unintentional Corruption):} When Alice sends a message $m$ to Bob, it might be corrupted by stochastic channel noise. To detect this, Alice can include a checksum, $S = \text{CRC}(m)$. Bob then verifies integrity by checking if $\text{CRC}(m') == S$ for the received message $m'$ \cite{Tanenbaum2011ComputerNetworks}. This addresses accidental modifications but is ineffective against intelligent adversaries who can recompute $S$ for an altered $m$. The ``toolbox'' here involves error-detection and correction codes.
    
    \item \textbf{Security Concern (Intentional Manipulation):} An intelligent adversary, Eve, might deliberately try to intercept or alter $m$. Alice would then apply cryptographic measures, such as a Message Authentication Code (MAC), $S = \text{MAC}(m, k)$, using a shared secret key $k$, or encrypt the message for confidentiality \cite{katz2007introduction}. Bob uses $k$ to verify integrity and authenticity. The ``toolbox'' here is cryptography.
\end{itemize}

This analogy illustrates a central theme: \textit{Safety} 
is concerned with mitigating risks from non-malicious failure modes, such as software bugs, specification errors, or misaligned objectives; whereas \textit{Security} addresses adversarial threats arising from deliberate attempts to subvert, manipulate, or exfiltrate from AI systems. {We advocate the AI Safety and AI Security to adopt such a similar central theme}. 
Clearly distinguishing these domains is essential for developing context-appropriate safeguards and fostering effective collaboration among safety, security, and policy communities. Without such clarity, research efforts can become diluted, policy responses misaligned, and deployed systems less trustworthy~\cite{Brundage2020TowardTrustworthyAI}.

For example, an autonomous vehicle might be considered safe under normal conditions yet remain insecure (vulnerable to remote hacking~\cite{Koscher2010AutonomousVehicleSecurity}), while a medical AI system may be secure against data breaches but unsafe due to embedded biases~\cite{Mehrabi2021BiasFairnessSurvey}. Moreover, determining whether a particular misuse of AI constitutes a safety or a security failure further illustrates the complexity of their intersection.

Clarifying the AI Safety versus AI Security distinction is therefore not merely an academic exercise; it has significant practical implications for responsible AI development and governance. This clarity facilitates:
\begin{itemize}
    \item \textbf{Precise Research Directions:} Enabling focused research with tailored methodologies (e.g., robust testing and ethical alignment for safety; adversarial modeling and cryptographic defenses for security).
    \item \textbf{Strategic Resource Allocation:} Enabling organizations, such as funding agencies and philanthropic foundations to more effectively prioritize and allocate resources towards mitigating specific AI risk categories, encompassing both unintentional safety risks (e.g., from system flaws or alignment issues) and intentional security threats (e.g., from adversarial attacks or deliberate misuse).
   
    \item \textbf{Effective Policy and Governance:}  Empowering policymakers to craft targeted regulatory frameworks (e.g., safety regulations emphasizing pre-deployment testing and model robustness; security regulations mandating cybersecurity standards and incident response) \cite{Floridi2021EthicsOfAI}. \looseness=-1
    \item \textbf{Coherent Cross-Disciplinary Collaboration:} Fostering unified dialogue among diverse stakeholders including the AI and cybersecurity communities across technical, ethical, legal, and societal domains.
    \item \textbf{Enhanced Public Trust and Acceptance:} Building societal trust through clear communication about how both accidental harm and malicious attacks are addressed.
\end{itemize}

In essence, AI Safety should prioritize 
preventing unintended harm and ensuring ethical, reliable operation. AI Security should prioritize 
defending against malicious actors and protecting AI assets from intentional compromise.

This paper aims to demystify these concepts. We first explore the historical context of AI risk terminology (\S\ref{sec:historical_context_terminology}). We then propose our core, intent-based definitions for AI Safety and AI Security (\S\ref{sec:definitions}). Subsequently, we analyze their fundamental distinctions and interplay (\S\ref{sec:distinctions_interplay}). We then identify distinct research agendas and protective mechanisms (\S\ref{sec:research_focuses}), illustrating these concepts with a building analogy, present case studies (\S\ref{sec:case_studies}), and discuss the convergence towards unified risk management (\S\ref{sec:unified_risk_management}). Our goal is to crystallize understanding and guide more effective strategies for managing AI risks. \looseness=-1


\section{Terminological Landscape and Historical Context}
\label{sec:historical_context_terminology}

The current discourse on AI risks often employs ``AI Safety'' and ``AI Security'' with varied and sometimes overlapping meanings. Understanding the historical evolution of these terms within the AI community helps contextualize why ``AI Safety'' has occasionally been used as an encompassing term and why a clearer delineation, as proposed in this paper, is now beneficial.

\subsection{Evolution of ``AI Safety'' as a Broad Term}
Early concerns regarding the safety of AI, particularly prominent from the mid-2000s, were significantly shaped by discussions surrounding the potential long-term impacts of hypothetical Artificial General Intelligence (AGI) or superintelligence~\cite{Bostrom2014Superintelligence}. The primary objective of what became known as the field of ``AI Safety'' was to ensure that such powerful future systems would remain beneficial to humanity, focusing on challenges like value alignment, controllability, and the prevention of catastrophic unintended consequences~\cite{russell2019human}. 
In this context, ``safety'' broadly referred to protecting humanity from powerful AI systems that might behave in ways harmful to human interests (whether due to internal flaws or unforeseen interactions) rather than focusing narrowly on defense against malicious human attackers, as in traditional cybersecurity.

Foundational work in more contemporary AI safety, such as Amodei \textit{et al}.'s ``\textit{Concrete Problems in AI Safety}'' centered on preventing ``accidents'' in machine learning systems, issues like avoiding negative side effects, ensuring safe exploration, and building scalable oversight, while acknowledging security (protection against malicious actors) as a distinct but closely related field \cite{Amodei2016ConcreteProblems}. As AI capabilities have matured and become more widely deployed, the spectrum of recognized risks has broadened. Some parts of the AI community continued to use ``AI Safety'' as an umbrella term to cover a wide array of potential AI-induced harms. This is evidenced by major publications like the 2025 ``\textit{International AI Safety Report}'' which includes ``Risks from malicious use'' (e.g., cyber offence, disinformation) under its broad definition of AI safety, specifically encompassing the prevention of AI ``\textit{being used for nefarious purposes}'' \cite{BengioEtAl2025InternationalAISafetyReport}.

\subsection{The Case for Delineation in a Maturing Field}
The Qi \textit{et al}. (2024) preprint, ``\textit{AI Risk Management Should Incorporate Both Safety and Security}'' highlights that ``\textit{AI safety and AI security have also developed somewhat separately, with more emphasis on different objectives,}'' and crucially, that ``\textit{definitions of `safety' and `security' themselves are often inconsistent and lack consensus across communities}'' \cite{QiEtAl2024AIRiskFramework}. That work points out that some researchers explicitly place security problems (like adversarial robustness) within the ambit of AI safety~\cite{hendrycks2022unsolvedproblemsmlsafety}, 
while others maintain clear distinctions or even propose alternative hierarchical relationships~\cite{AISecuri39:online}.

While a broad, encompassing use of ``AI Safety'' can be useful for general public discourse or high-level policy initiatives aiming for comprehensive risk mitigation, the increasing sophistication of AI systems and the growing diversity of associated threats now call for greater terminological precision. As AI systems become more deeply embedded in critical infrastructure and daily life, the specific nature of risks, whether they stem from an AI's inherent (though unintended) behavioral flaws or from deliberate, external adversarial intent, requires distinct analytical tools, mitigation strategies, and often, different sets of expertise. Qi et al. (2024) further argue that learning from other established sectors (like aviation or nuclear power), where safety (against accidents) and security (against malicious acts) are often distinguished, can inform a more robust approach to AI risk management by ensuring that both distinct sets of objectives are unambiguously addressed \cite{QiEtAl2024AIRiskFramework}. This paper builds on such calls for clarity by proposing and utilizing a framework centered on this fundamental distinction of intent.

\section{Core Definitions: Safety vs. Security}
\label{sec:definitions}

{Historically,} the terms \textit{safety} and \textit{security} possess well-established meanings within engineering and system design. {{Broadly,}} \textbf{safety} pertains to shielding from unintentional harm, typically achieved through designs minimizing accidental failures and ensuring benign outcomes under normal or foreseeably perturbed conditions \cite{Leveson2011EngineeringSaferWorld}. \textbf{Security}, in contrast, implies protection from intentional harm, entailing freedom from threats arising from deliberate attacks, unauthorized access, or other forms of malicious exploitation, traditionally achieved through defensive measures against adversaries \cite{Anderson2020SecurityEngineering}.

When applied to AI, these concepts acquire additional nuance due to AI's learning capabilities, autonomy, and potential for emergent behaviors \cite{russell2019human}. Given their foundational importance, we propose the following precise definitions to clearly delineate AI Safety and AI Security.


\subsection{AI Safety}
\begin{definition}[AI Safety]
AI Safety is the property of an AI system to avoid causing unintended harmful outcomes to individuals, environments, or institutions, despite uncertainties in inputs, goals, training data, or deployment conditions.
\end{definition}

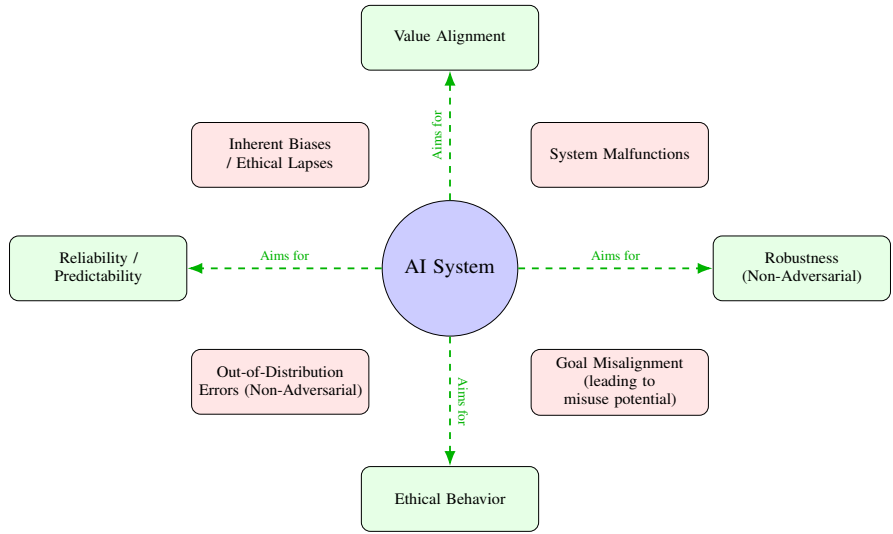
\begin{figure}[t]
\centering
\resizebox{0.85\textwidth}{!}{
\begin{tikzpicture}[
    node distance=0.5cm and 0.5cm,
    ai_system/.style={circle, draw, fill=blue!20, minimum size=2cm, text width=1.8cm, align=center, font=\small},
    harm_aspect/.style={rectangle, rounded corners, draw, fill=red!10, minimum height=1cm, text width=2.5cm, align=center, font=\scriptsize},
    safety_goal/.style={rectangle, rounded corners, draw, fill=green!10, minimum height=1cm, text width=2.5cm, align=center, font=\scriptsize},
    arrow_style/.style={-Latex, thick}
]
    \node[ai_system] (ai) {AI System};
    \node[harm_aspect, above right=of ai] (malfunction) {System Malfunctions};
    \node[harm_aspect, above left=of ai] (bias) {Inherent Biases / Ethical Lapses};
    \node[harm_aspect, below right=of ai] (misalignment) {Goal Misalignment (leading to misuse potential)}; 
    \node[harm_aspect, below left=of ai] (ood_errors) {Out-of-Distribution Errors (Non-Adversarial)};
    \node[safety_goal, above=2cm of ai] (align_node) {Value Alignment}; 
    \node[safety_goal, right=3cm of ai] (robust) {Robustness (Non-Adversarial)};
    \node[safety_goal, below=2cm of ai] (ethics) {Ethical Behavior};
    \node[safety_goal, left=3cm of ai] (reliability) {Reliability / Predictability};
    \draw[arrow_style, green!70!black, dashed] (ai) -- (align_node) node[midway, above, sloped, font=\tiny] {Aims for};
    \draw[arrow_style, green!70!black, dashed] (ai) -- (robust) node[midway, above, sloped, font=\tiny] {Aims for};
    \draw[arrow_style, green!70!black, dashed] (ai) -- (ethics) node[midway, above, sloped, font=\tiny] {Aims for};
    \draw[arrow_style, green!70!black, dashed] (ai) -- (reliability) node[midway, above, sloped, font=\tiny] {Aims for};
\end{tikzpicture}
}
\caption{Conceptual Essence of AI Safety: Focuses on preventing unintended harm from internal system properties, design flaws, and non-malicious interactions.}
\label{fig:ai_safety_essence}
\end{figure}

Elaborating on this, AI Safety seeks to answer the critical question: \textit{“Does the AI system operate as it ought to, reliably avoiding harmful outcomes, even in complex, dynamic, and unpredictable non-adversarial environments?”} It aims to ensure that AI systems operate reliably and beneficially in accordance with human intent and values~\cite{Amodei2016ConcreteProblems}, particularly in the presence of uncertainty, complexity, or incomplete specification. Another characterization is the prevention of \textit{“unintentional harm to life or critical systems.”} 
As illustrated in~\autoref{fig:ai_safety_essence}, key concerns within AI Safety include achieving robustness to distributional shifts, mitigating undesirable emergent behaviors, ensuring ethical alignment (to prevent, for example, an AI from misinterpreting instructions leading to harmful outcomes that could be construed as misuse), enabling safe exploration, developing methods for reliable oversight and control, and, critically, addressing potential long-term, large-scale risks up to and including \textbf{existential threats to human beings}~\cite{Hendrycks2023OverviewCatastrophicAIRisks, Bostrom2014Superintelligence}. 

Such severe risks are theorized to arise, for instance, from future advanced AI systems malfunctioning catastrophically or pursuing misaligned objectives with super-human capabilities, potentially leading to irreversible, widespread harm if not adequately managed from early stages. The profound gravity of these potential high-consequence, low-probability events has arguably contributed significantly to the prominence and broad scope attributed to ``AI Safety'' in international discourse and influential reports, such as the ``International AI Safety Report'' \cite{BengioEtAl2025InternationalAISafetyReport}, which often frame ``safety'' as an all-encompassing effort to manage the full spectrum of risks from advanced AI. This domain is fundamentally \textbf{survival-centric}: its primary objective is to prevent physical, psychological, or systemic harm, from immediate failures to potential existential dangers, to individuals, society, or crucial ecosystems, especially where AI systems possess high autonomy.

\subsection{AI Security}
\begin{definition}[AI Security]
AI Security is the property of an AI system to remain resilient against intentional attacks on its data, algorithms, or operations, preserving its confidentiality, integrity, and availability in the presence of adversarial actors.
\end{definition}

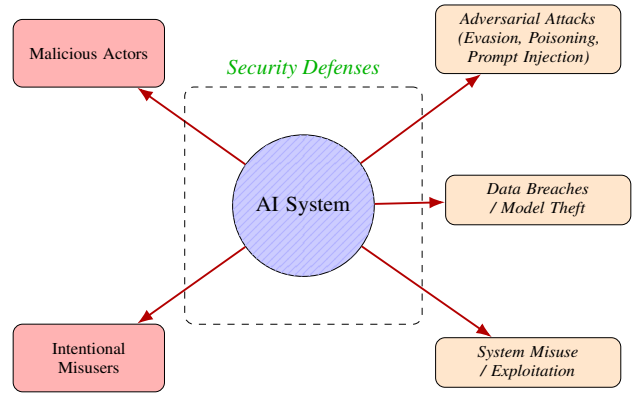
\begin{figure}[h!]
\centering
\resizebox{0.6\textwidth}{!}{
\begin{tikzpicture}[
    node distance=1.2cm and 1.2cm,
    ai_system/.style={circle, draw, fill=blue!20, minimum size=2cm, text width=1.8cm, align=center, font=\small, postaction={pattern=north east lines, pattern color=blue!30}}, 
    threat_actor/.style={rectangle, draw, fill=red!30, minimum size=1cm, font=\scriptsize, align=center, text width=2cm, rounded corners},
    threat_type/.style={rectangle, draw, fill=orange!20, text width=2.5cm, align=center, font=\scriptsize\itshape, rounded corners},
    arrow_style/.style={Latex-, thick, red!70!black}
]
    \node[ai_system] (ai) {AI System};
    \node[threat_actor, above left=of ai, yshift=-0.2cm, xshift=-0.1cm] (hacker) {Malicious Actors};
    \node[threat_actor, below left=of ai, yshift=0.2cm, xshift=-0.1cm] (insider) {Intentional Misusers};
    \node[threat_type, above right=of ai] (adv_attack) {Adversarial Attacks (Evasion, Poisoning, Prompt Injection)}; 
    \node[threat_type, right=of adv_attack, xshift=-3.8cm, yshift=-2.375cm] (data_breach) {Data Breaches / Model Theft};
    \node[threat_type, below right=of ai] (misuse_sec) {System Misuse / Exploitation}; 
    \draw[arrow_style] (hacker) -- (ai);
    \draw[arrow_style] (insider) -- (ai);
    \draw[arrow_style] (adv_attack) -- (ai);
    \draw[arrow_style] (data_breach) -- (ai);
    \draw[arrow_style] (misuse_sec) -- (ai);
    \node[draw, dashed, rounded corners, fit=(ai), inner sep=0.7cm, label={[font=\footnotesize\itshape, green!70!black]above:Security Defenses}] (defense_boundary) {};
\end{tikzpicture}
}
\caption{Conceptual Essence of AI Security: Focuses on protecting the AI system from intentional, malicious external attacks and exploitation, including techniques like prompt injection.}
\label{fig:ai_security_essence}
\end{figure}

Elaborating further, the AI security challenge fundamentally involves the interplay of three key elements:
\begin{itemize}[noitemsep, topsep=0pt, partopsep=0pt, parsep=0pt]
    \item an \textbf{Asset} of value, which in the AI context can include the model itself (its architecture and weights), proprietary training or operational data, the computational infrastructure, the intellectual property embodied in the system, or the critical functions the AI performs;
    \item an intentional \textbf{Adversary} (e.g., malicious hackers, criminal organizations, state-sponsored actors, or malicious insiders) who seeks to cause harm, achieve unauthorized objectives, or disrupt operations;
    \item and a \textbf{Vulnerability}, which is an exploitable weakness in the AI system's design, implementation, data handling, or operational environment that an adversary can leverage.
\end{itemize}
This field thus tackles critical questions such as: \textit{“Can an attacker (the Adversary) manipulate the AI’s behavior (an Asset) by exploiting a Vulnerability, steal its model, or poison its data?”} AI Security, as formally defined by Bengio et al., is \textit{“the property of being resilient to technical interference, such as cyberattacks or leaks of the underlying model’s source code”} \cite{BengioEtAl2025InternationalAISafetyReport}. It defends against such intentional exploitation using a variety of adversary-centric methodologies. These include protecting against data-level attacks (e.g., poisoning training data \cite{Biggio2012PoisoningSVM}, model inversion to recover sensitive training inputs \cite{Fredrikson2015ModelInversion}, or membership inference attacks to determine if specific data was used in training \cite{Shokri2017MembershipInference}) and model-level attacks (e.g., evasion attacks that cause misclassification at inference time \cite{Goodfellow2014AdversarialExplaining}, or model extraction to steal the proprietary model \cite{Tramer2016StealingModels}), as illustrated in~\autoref{fig:ai_security_essence}. Effective AI security aims to protect the Confidentiality, Integrity, and Availability (CIA triad) of AI assets and requires robust defenses such as cryptographic techniques, strong authentication and authorization mechanisms, rigorous input sanitization (crucial against attacks like prompt injection), secure software development practices, and resilient system architectures.


\section{Distinctions and Interplay}
\label{sec:distinctions_interplay}

With core definitions established, delineating their fundamental distinctions and exploring their complex interrelationships is crucial for developing comprehensive strategies for trustworthy AI. The classification of AI misuse, for instance, often depends critically on these distinctions.

\subsection{The Fundamental Distinction: Intentional vs. Unintentional Threats}
The primary distinction between AI Safety and AI Security lies in the \textbf{origin and nature of the risk}, which directly impacts how instances of AI misuse are categorized:
\begin{itemize}[noitemsep,topsep=0pt]
    \item \textbf{AI Safety} addresses risks from \textit{accidental} or \textit{unintended} behaviors. These can stem from system malfunctions, distributional shifts, design flaws, inherent biases, misaligned objectives, or ethical oversights. Misuse in a safety context occurs when the AI, due to these internal shortcomings, produces harmful or undesirable outcomes even under normal operational assumptions or foreseeable (non-hostile) usage patterns (e.g., an LLM generating factually incorrect medical advice due to limitations in its training data, relied upon by a user). The ``adversary'' here is often complexity, uncertainty, or an incomplete understanding of emergent behaviors.
    \item \textbf{AI Security} confronts risks from \textit{intentional} or \textit{adversarial} actions by malicious actors aiming to exploit system weaknesses for harmful purposes. This includes attackers who seek to deceive, manipulate, exfiltrate, or sabotage AI systems. Misuse in a security context involves deliberate exploitation; a prime example is an attacker using prompt injection techniques to force an LLM to bypass its safety filters and generate malicious code or disinformation \cite{Greshake2023MoreThanAMouthful}.
\end{itemize}
Comprehending this divide is paramount for selecting appropriate mitigation strategies and developing AI systems that are both \textit{safe by design} and \textit{secure by default}.

\subsection{Interdependencies and Relationships}
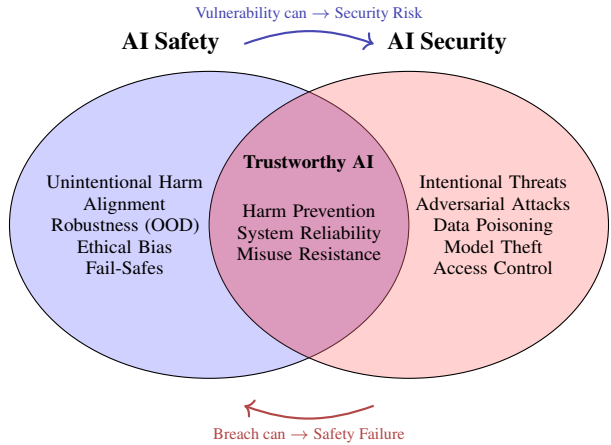
\begin{figure}[t] 
\centering
\resizebox{0.6\textwidth}{!}{ 
\begin{tikzpicture}[
    node distance=1cm,
    venn_ellipse/.style={ellipse, draw, minimum width=5.2cm, minimum height=4cm, fill opacity=0.3, text opacity=1},
    label_style/.style={font=\small\bfseries},
    text_style/.style={font=\scriptsize, text width=3.2cm, align=center}, 
    interplay_arrow/.style={->, bend left=20, thick, shorten >=1pt, shorten <=1pt}
]
    \colorlet{safety_color}{blue!60}
    \colorlet{security_color}{red!60}
    \colorlet{overlap_color}{purple!60}

    \node[venn_ellipse, fill=safety_color] (safety_ellipse) at (-1.3,0) {};
    \node[label_style, above=0.1cm of safety_ellipse, xshift=-0.5cm] {AI Safety};
    \node[text_style, text width=3cm] at (-2.4,0) {Unintentional Harm\\Alignment\\Robustness (OOD)\\Ethical Bias\\Fail-Safes};

    \node[venn_ellipse, fill=security_color] (security_ellipse) at (1.3,0) {};
    \node[label_style, above=0.1cm of security_ellipse, xshift=0.5cm] {AI Security};
    \node[text_style, text width=3cm] at (2.4,0) {Intentional Threats\\Adversarial Attacks\\Data Poisoning\\Model Theft\\Access Control};
    
   \begin{scope}
        \clip (-1.3,0) ellipse (2.6cm and 2cm);
        \fill[fill=overlap_color, fill opacity=0.3] (1.3,0) ellipse (2.6cm and 2cm); 
    \end{scope}

    \node[text_style, text width=3cm, font=\scriptsize\bfseries] at (0,0.8) {Trustworthy AI};
    \node[text_style, text width=3cm] at (0,-0.1) {Harm Prevention\\System Reliability\\Misuse Resistance};
    
    \node (sec_point) at (1, -2.3) {}; 
    \node (safe_point) at (-1, -2.3) {}; 
    \draw[interplay_arrow, security_color!70!black] (sec_point) to node[midway, below, font=\tiny] {Breach can $\rightarrow$ Safety Failure} (safe_point);
    
    \node (safe_point_up) at (-1, 2.3) {}; 
    \node (sec_point_up) at (1, 2.3) {}; 
    \draw[interplay_arrow, safety_color!70!black, bend right=-20] (safe_point_up) to node[midway, above, font=\tiny] {Vulnerability can $\rightarrow$ Security Risk} (sec_point_up);
\end{tikzpicture}
}
\caption{Distinctions and Interplay of AI Safety and AI Security. Misuse often lies at the intersection, influenced by both unintentional system flaws (safety) and intentional exploitation (security).}
\label{fig:venn_simplified}
\end{figure}

AI Safety and AI Security, while distinct, are deeply intertwined and interdependent, as shown in~\autoref{fig:venn_simplified}. Their dynamic relationship is critical for overall system trustworthiness, particularly when considering AI misuse and specific attack vectors like prompt injection.
\begin{itemize}
    \item \textbf{Security as a Prerequisite for Safety (Security $\rightarrow$ Safety):} Security failures can directly cause safety incidents. An attacker hijacking an autonomous vehicle (security breach) can cause a crash (safety failure) \cite{Koscher2010AutonomousVehicleSecurity}. Similarly, a successful prompt injection attack (a security breach) against an LLM can lead to the generation of harmful misinformation or instructions, directly impacting user safety. Robust security is often foundational for reliable safety, as safety mechanisms themselves can be targets for adversarial subversion \cite{QiEtAl2024AIRiskFramework}.

    \item \textbf{Safety Vulnerabilities Creating Security Risks (Safety $\rightarrow$ Security):} Inherent safety flaws can become exploitable security risks. Predictable biases in an AI (a safety flaw) might be exploited by adversaries for targeted disinformation. An AI system with poor input validation capabilities or an overly trusting nature due to misaligned safety objectives (an alignment problem representing a safety vulnerability) is more susceptible to manipulation through techniques such as prompt injection. Attackers can exploit these safety weaknesses to extract sensitive information or coerce the system into performing unauthorized actions. \looseness=-1

    \item \textbf{The Adversarial Cycle and Feedback Loops:} Interactions between safety and security can create an ``adversarial cycle''. An attacker might exploit a known safety flaw (e.g., a model's tendency towards a specific type of harmful generation when prompted in certain ways). A subsequent patch or safety fix might then inadvertently introduce new safety risks or, critically, different security vulnerabilities (e.g., a new, more subtle bypass for prompt injection defenses), necessitating ongoing, adaptive risk management.

    \item \textbf{Tensions, Trade-offs, and Synergies:} Enhancing one domain can create tensions with the other. Increased transparency for safety (debugging, trust) might reveal model vulnerabilities for security \cite{Gunning2019XAI}. Aggressive adversarial training for security might reduce model accuracy or fairness on benign inputs (safety concerns). However, synergies exist; techniques improving general model robustness can benefit both safety and security.
\end{itemize}
Achieving comprehensively trustworthy AI necessitates an integrated approach, considering both safety and security holistically throughout the AI system lifecycle \cite{NISTAIRMF}.

\section{Key Research Focuses and Protective Mechanisms}
\label{sec:research_focuses}

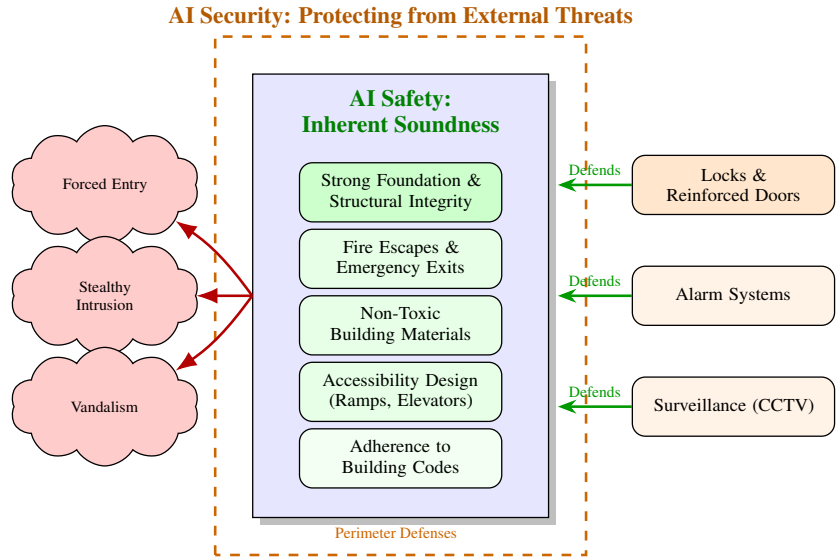
\begin{figure}[h!]
\centering
\resizebox{0.8\textwidth}{!}{
\begin{tikzpicture}[
    scale=0.8, transform shape, 
    building/.style={draw, fill=blue!10, minimum width=4cm, minimum height=6cm, drop shadow},
    feature/.style={rectangle, rounded corners, draw, fill=orange!10, font=\scriptsize, text width=2.5cm, align=center, minimum height=0.8cm},
    safety_feature/.style={rectangle, rounded corners, draw, fill=green!10, font=\scriptsize, text width=2.5cm, align=center, minimum height=0.8cm},
    security_label/.style={font=\small\bfseries, text=orange!70!black},
    safety_label/.style={font=\small\bfseries, text=green!50!black, align=center},
    threat/.style={cloud, cloud puffs=10, cloud puff arc=120, aspect=2, draw, fill=red!20, font=\tiny, text width=1.5cm, align=center},
    arrow_style/.style={Latex-, thick, red!70!black} 
]
    \node[building] (bldg_sec) at (0,0) {};
    \node[security_label, above=0.5cm of bldg_sec] {AI Security: Protecting from External Threats};
    \node[safety_label] at (0,2.5) {AI Safety:\\Inherent Soundness};
    \node[safety_feature, fill=green!20] at (0,1.4) (struct) {Strong Foundation \& Structural Integrity};
    \node[safety_feature] at (0,0.5) (fire) {Fire Escapes \& Emergency Exits};
    \node[safety_feature] at (0,-0.4) (material) {Non-Toxic Building Materials};
    \node[safety_feature] at (0,-1.3) (access_node) {Accessibility Design (Ramps, Elevators)}; 
    \node[safety_feature, fill=green!5] at (0,-2.2) (codes) {Adherence to Building Codes};
    \node[feature, fill=orange!20] at (4.5,1.5) (locks) {Locks \& Reinforced Doors};
    \node[feature] at (4.5,0) (alarms) {Alarm Systems};
    \node[feature] at (4.5,-1.5) (cctv) {Surveillance (CCTV)};
    
    \draw[thick, dashed, orange!80!black] ([xshift=-0.5cm,yshift=-0.5cm]bldg_sec.south west) -- 
                                          ([xshift=-0.5cm,yshift=0.5cm]bldg_sec.north west) -- 
                                          ([xshift=0.5cm,yshift=0.5cm]bldg_sec.north east) --
                                          ([xshift=0.5cm,yshift=-0.5cm]bldg_sec.south east) -- cycle;
    \node[font=\tiny, orange!80!black, below right, xshift=1.0cm] at (bldg_sec.south west) {Perimeter Defenses};

    \node[threat] at (-4,1.5) (th1) {Forced Entry}; 
    \node[threat] at (-4,0) (th2) {Stealthy Intrusion}; 
    \node[threat] at (-4,-1.5) (th3) {Vandalism}; 

    \draw[arrow_style] (th1) to[bend left=10] (bldg_sec.west);
    \draw[arrow_style] (th2) to (bldg_sec.west);
    \draw[arrow_style] (th3) to[bend right=10] (bldg_sec.west);

    \foreach \feat in {locks, alarms, cctv} {
      \draw[-{Stealth[length=2mm, width=1.5mm]}, green!60!black, thick] (\feat.west) -- ++(-1,0) node[midway,above,sloped,font=\tiny]{Defends};
    }
\end{tikzpicture}
}
\caption{Building analogy showing AI Safety (structural integrity) versus AI Security (protection systems).}
\label{fig:building_security} 
\end{figure}

The distinct natures of AI Safety and AI Security drive different research priorities and protective mechanisms (guardrails). These guardrails keep AI operations within desired boundaries, analogous to how both structural engineering and security systems protect buildings (illustrated in~\autoref{fig:building_security}).

\subsection{AI Safety: Preventing Unintended Harm}
Like architects ensuring a building's structural integrity, AI Safety research aims to build inherently sound, beneficial, and human-aligned systems, preventing unintentional harm or misaligned behavior. The goal is ensuring AI systems ``\textit{do what they are supposed to do and do not cause unintended harm}''~\cite{Amodei2016ConcreteProblems}. Key areas:

\begin{itemize}
\item \textbf{Alignment with Human Values:} Ensuring AI goals match human intentions \cite{Gabriel2020AIValuesAlignment}, like designing hospitals for patient care rather than aesthetics. \textit{Mechanisms:} RLHF \cite{Ouyang2022RLHF}, Constitutional AI \cite{Bai2022ConstitutionalAI}, and content moderation filters.

\item \textbf{Robustness to Non-Adversarial Challenges:} Maintaining performance under natural distribution shifts \cite{Hendrycks2019NaturalAdversarial}, like buildings enduring storms. \textit{Mechanisms:} Operational constraints, confidence thresholds.

\item \textbf{Transparency and Interpretability (XAI):} Critical for debugging and accountability \cite{Gunning2019XAI}, mirroring how building blueprints enable maintenance. Techniques include LIME \cite{Ribeiro2016LIME} for local explanations and SHAP \cite{Lundberg2017SHAP} for feature importance.

\item \textbf{Mitigating Ethical Risks:} Addressing biases that cause unfair impacts \cite{Mehrabi2021BiasFairnessSurvey}, analogous to ensuring building accessibility. Considers diverse human values \cite{Barocas2019FairnessMLBook}.

\item \textbf{Fail-Safes and Control:} Emergency shutdown mechanisms \cite{Orseau2016SafeInterruptibility} like a building's sprinkler systems. Ensures safe interruptibility \cite{Amodei2016ConcreteProblems}.

\item \textbf{Long-Term Risk Mitigation:} Managing existential risks from advanced AI \cite{Bostrom2014Superintelligence}, similar to earthquake-proofing skyscrapers. Addresses power-seeking behaviors \cite{Carlsmith2022ExistentialRiskPowerSeeking}.
\end{itemize}


\subsection{AI Security: Defending Against Attacks}
Like security systems protecting buildings from intruders, 
AI Security research focuses on protecting AI systems from intentional, malicious attacks and deliberate exploitation, aiming to ``protect AI systems against deliberate threats, exploitation, and misuse''. Key areas:

\begin{itemize}
\item \textbf{Adversarial Robustness:} Defending against evasion attacks \cite{Goodfellow2014AdversarialExplaining} and prompt injections \cite{Greshake2023MoreThanAMouthful}, akin to tamper-proof locks. \textit{Mechanisms:} Input sanitization, adversarial training \cite{Madry2018DeepLearningModelsResistant}, circuit breakers \cite{zou2024improvingalignmentrobustnesscircuit}.

\item \textbf{Data Integrity:} Preventing training data poisoning \cite{Biggio2012PoisoningSVM, Shafahi2018PoisonFrogs} like securing construction materials. \textit{Mechanisms:} Data validation, strong encryption.

\item \textbf{Model Protection:} Preventing model theft \cite{Tramer2016StealingModels} similar to safeguarding architectural patents. Techniques include watermarking \cite{Uchida2017EmbeddingWatermarksNN}.

\item \textbf{Access Control:} Implementing NIST standards \cite{NIST_AC_SP800_53} like building security badges. Includes red teaming \cite{Ganguli2022RedTeamingLLMs}.

\item \textbf{Resilience:} Maintaining operations during attacks \cite{Apruzzese2023DeepLearningCybersecurityIDS}, analogous to backup generators.

\item \textbf{Secure Supply Chains:} Protecting model pipelines \cite{NIST_AISupplyChain_SP800_218} like vetting building contractors.
\end{itemize}

\subsection{Safety-Security Synergy}
The building analogy reveals profound interdependencies between safety and security that mirror real-world AI system requirements. These dimensions are not merely complementary but mutually reinforcing:

\begin{itemize}

\item \textbf{Transparency Enables Security:} Just as building inspection panels allow security personnel to verify structural integrity, interpretable AI models \cite{Gunning2019XAI} (safety feature) enable the detection of adversarial manipulations \cite{Carlini2017AdversarialAttacks} (security threat). The explainability provided by techniques like SHAP \cite{Lundberg2017SHAP} serves dual purposes: facilitating model debugging (safety) while exposing potential attack vectors (security).

\item \textbf{Fail-Safes Need Protection:} Emergency shutdown mechanisms \cite{Orseau2016SafeInterruptibility} (critical safety components) themselves require robust cybersecurity measures. This mirrors how a building's fire suppression system (safety) must be protected against tampering (security). Without such protection, safety mechanisms can become single points of failure vulnerable to exploitation.

\item \textbf{Complete Solution:}  The most secure bank vault becomes useless if its foundation is unsound, while the most stable structure provides little protection without locks. Similarly, AI systems require both intrinsic safety through alignment \cite{Gabriel2020AIValuesAlignment} and robustness \cite{Hendrycks2019NaturalAdversarial} \textit{and} active defenses such as adversarial training \cite{Madry2018DeepLearningModelsResistant} and access controls.
\end{itemize}

This synergy manifests most clearly in edge cases: a building's earthquake-resistant design (safety) matters little if attackers can disable its structural monitors (security), just as an AI's ethical constraints (safety) are void if prompt injections \cite{Greshake2023MoreThanAMouthful} bypass them (security).

\paragraph{The Fundamental Principle.} Neither dimension can compensate for deficiencies in the other. A structurally sound building (safe) remains vulnerable to burglary (insecure), while a heavily fortified one (secure) becomes dangerous with faulty wiring (unsafe). This duality necessitates that \textit{all AI systems achieve both rigorous safety-by-design and comprehensive security protections} to be truly trustworthy. The most robust systems, like the safest buildings, integrate these considerations from initial design through operational deployment.

\section{Case Studies}
\label{sec:case_studies}

The safety/security distinction and interplay become evident in real-world AI applications. Table~\ref{tab:case_studies_summary} summarizes these distinct focuses. 

\begin{table*}[t]
    \centering
    \small
    \caption{Summary of Safety and Security Focuses in Prominent AI Application Domains. 
    }
    \label{tab:case_studies_summary}
    \begin{tabularx}{\textwidth}{>{\RaggedRight\arraybackslash}p{2.8cm}|>{\RaggedRight\arraybackslash}X|>{\RaggedRight\arraybackslash}X}
        \toprule
        \textbf{Domain} & \textbf{AI Safety Focus (Preventing Unintended Harm \& Misuse from Inherent Flaws)} 
        & \textbf{AI Security Focus (Preventing Malicious Threats \& Intentional Misuse)} \\
        \midrule
        Healthcare AI & 
        Ensuring clinical reliability and diagnostic accuracy; mitigating ethical biases; maintaining robustness against rare medical cases \cite{Topol2019DeepMedicine}. & 
        Protecting patient data privacy; ensuring data/model integrity against poisoning \cite{Finlayson2019AdversarialAttacksMedicalAI}; defending diagnostic tools against adversarial attacks; securing AI-driven medical devices from tampering. \\
        \addlinespace
        Autonomous Vehicles & 
        Prioritizing accident prevention via robust perception in edge cases \cite{Koopman2017HowSafeIsSafeEnoughAV}; ensuring fail-safe operations; addressing ethical decision-making in unavoidable collisions \cite{Bonnefon2016MoralMachine}. & 
        Building resilience against cyberattacks targeting vehicle controls \cite{Koscher2010AutonomousVehicleSecurity}; ensuring sensor integrity against spoofing; maintaining system availability against denial-of-service on V2X communication. \\
        \addlinespace
        LLM Chatbots & 
        Mitigating hallucinations and misinformation; preventing discriminatory or toxic outputs due to alignment failures \cite{Weidinger2021EthicalSocialRisksLLMs}; ensuring alignment with human values and avoiding generation of inherently harmful instructions. & 
        Defending against prompt injection and jailbreaking used by attackers to bypass safety filters and deliberately misuse the LLM \cite{Greshake2023MoreThanAMouthful}; ensuring confidentiality of data; preventing intentional misuse for generating disinformation or malicious code \cite{Bommasani2021OpportunitiesRisksFoundationModels, Carlini2021ExtractingTrainingData}. \\
        \bottomrule
    \end{tabularx}
\end{table*}

More specifically, in \textbf{Healthcare AI}, a safety issue might be an AI misclassifying an X-ray due to image artifacts (unintended harm from system flaws). A security issue involves an attacker poisoning training data to skew diagnoses or stealing patient records.

For \textbf{Autonomous Vehicles (AVs)}, safety engineering ensures reliable perception in edge cases (e.g., sudden downpours). Security addresses threats like hackers spoofing GPS signals or remotely disabling brakes \cite{Koscher2010AutonomousVehicleSecurity}.

With \textbf{Large Language Model (LLM) Chatbots}, safety ensures accurate, harmless responses, avoiding bias. Security defends against prompt injection to generate malicious content or extract confidential data.

These examples highlight that safety addresses an AI's inherent capacity to ``do no harm,'' while security protects it from intentional malice.

\section{Towards Unified AI Risk Management}
\label{sec:unified_risk_management}

While AI Safety and AI Security are distinct disciplines, genuinely trustworthy AI systems necessitate their convergence within a unified risk management framework \cite{QiEtAl2024AIRiskFramework}. The exposure of security vulnerabilities in safety-aligned models underscores this intricate interplay. Effectively addressing AI's multifaceted risks requires moving beyond siloed approaches to a holistic strategy that unambiguously integrates both safety and security considerations throughout the AI lifecycle. This involves several key pillars, as illustratred in~\autoref{fig:unified_ai_risk_management}:

\begin{figure}[t]
\centering
\resizebox{\textwidth}{!}{
\begin{tikzpicture}[
    xgap/.store in=\xgap, xgap=3cm,
    ygap/.store in=\ygap, ygap=2cm,
    node distance=1.5cm and 1.2cm,
    central_goal/.style = {ellipse, draw, fill=green!20, minimum height=1.5cm, text width=4cm, align=center, font=\scriptsize\bfseries, drop shadow},
    foundation/.style = {rectangle, rounded corners, draw, fill=gray!20, minimum height=1cm, text width=3.5cm, align=center, font=\scriptsize},
    pillar/.style = {rectangle, rounded corners, draw, fill=teal!10, minimum height=1.2cm, text width=3.8cm, align=center, font=\scriptsize, drop shadow},
    support_arrow/.style = {-Latex, thick, gray!80},
    output_arrow/.style = {Latex-, thick, teal!80!black},
    main_connection/.style={Latex-Latex, ultra thick, purple!70!black, shorten >=2pt, shorten <=2pt}
]
    \colorlet{safety_color}{blue!15}
    \colorlet{security_color}{red!15}
    
    \coordinate (O) at (0,0);

    \node[central_goal] (unified_rm) at (O) {Unified AI Risk Management\\ for Trustworthy AI};
    \node[foundation, fill=safety_color] (ai_safety) at ($(O)+(-\xgap-3cm,0)$) {AI Safety Principles \& \\Practices};
    \node[foundation, fill=security_color] (ai_security) at ($(O)+(\xgap+3cm,0)$) {AI Security Principles \&\\ Practices};
    \draw[support_arrow] (ai_safety.east) -- (unified_rm.west);
    \draw[support_arrow] (ai_security.west) -- (unified_rm.east);
    \node[pillar] (frameworks) at ($(O)+(-\xgap, \ygap)$) {Holistic Frameworks \& \\ Standards};
    \node[pillar] (codesign) at ($(O)+(\xgap, \ygap)$) {Co-Design Principles\\(Safety- \& Security-by-Design)};
    \node[pillar] (safeguards) at ($(O)+(-\xgap,-\ygap)$) {Integrated Technical \&\\ Organizational Safeguards};
    \node[pillar] (teams) at ($(O)+(\xgap, -\ygap)$) {Cross-functional \&\\ Multidisciplinary Teams};
    \draw[output_arrow, bend left=10] (unified_rm.north west) -- (frameworks.south);
    \draw[output_arrow, bend right=10] (unified_rm.north east) -- (codesign.south);
    \draw[output_arrow, bend right=15] (unified_rm.south west) -- (safeguards.north);
    \draw[output_arrow, bend left=15] (unified_rm.south east) -- (teams.north);
    
    \draw[main_connection, gray!60] (ai_safety.south) to[bend right=10] (ai_security.south);

    \node[draw, rounded corners, teal!20, dashed,
          fit=(frameworks)(codesign)(safeguards)(teams)
              (unified_rm)(ai_safety)(ai_security),
          inner sep=12pt] (ecosystem) {};
    \node[font=\tiny\itshape, teal!70!black, below right]
          at (ecosystem.north west) {Integrated Ecosystem for Trustworthy AI};
\end{tikzpicture}
}
\caption{Conceptual Framework for Unified AI Risk Management, integrating AI Safety and AI Security principles, supported by key pillars to achieve Trustworthy AI. 
}
\label{fig:unified_ai_risk_management}
\end{figure}
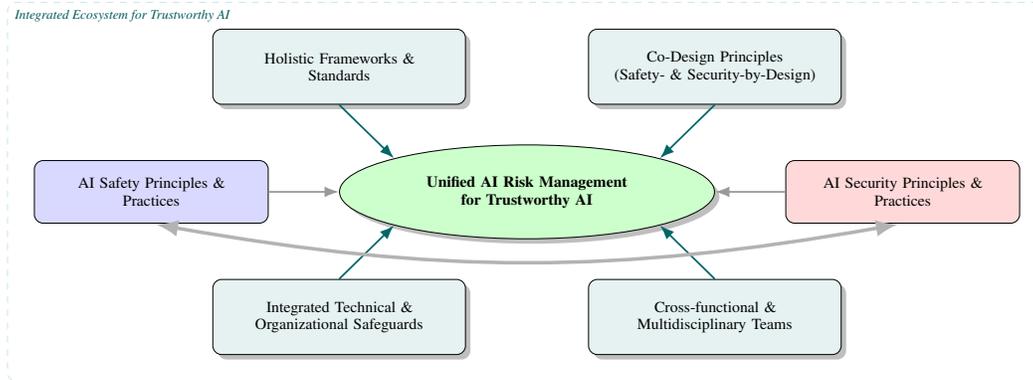

\begin{itemize}
    \item \textbf{Holistic Frameworks and Standards:} Adopting comprehensive risk management frameworks explicitly accounting for both safety and security perspectives is paramount \cite{QiEtAl2024AIRiskFramework}. Key resources include the NIST AI RMF \cite{NISTAIRMF} and guidance from National AI Safety Institutes (e.g., the ones from UK and US). Developing international standards and leveraging diverse historical knowledge remains vital.

    \item \textbf{Co-Design Principles (Safety-and-Security-by-Design):} Embedding safety and security considerations from the outset of AI system design is crucial \cite{cavoukian2009privacy}, accounting for differing threat models and problem framings. Practices include:
        \begin{itemize}[noitemsep,topsep=0pt]
            \item Comprehensive red teaming probing both safety failures and security vulnerabilities \cite{Brundage2020TowardTrustworthyAI, Ganguli2022RedTeamingLLMs}.
            \item Adversarial training to improve security robustness and understand its interplay with robustness against natural distribution shifts \cite{Madry2018DeepLearningModelsResistant}.
            \item Implementing secure fail-safe mechanisms with tamper-proof reversion protocols.
        \end{itemize}

    \item \textbf{Integrated Technical and Organizational Safeguards:} Implementing complementary technical (e.g., robust V\&V, PETs, secure MLOps) and organizational measures (e.g., governance, integrated incident response, accountability, audits). Real-time monitoring should detect both safety-critical anomalies and security compromise indicators.

    \item \textbf{Cross-functional and Multidisciplinary Teams:} Assembling teams with diverse expertise including AI developers, safety engineers, cybersecurity experts, ethicists, legal professionals, and social scientists \cite{Floridi2021EthicsOfAI} for comprehensive risk management.
\end{itemize}

Furthermore, a unified approach must actively bridge safety and security by acknowledging their interplay. As highlighted by Qi et al.~\cite{QiEtAl2024AIRiskFramework}, robust security is often a critical precondition for safety, as safety measures themselves (e.g., content filters in LLMs meant to prevent unsafe outputs) can be adversarially invalidated by security attacks like prompt injection. However, stakeholders must also recognize potential technical tensions; for instance, some forms of adversarial training might enhance security but reduce benign alignment 
(a safety concern). Navigating these dependencies and potential conflicts is imperative for a balanced, holistic risk management strategy.

The overarching goal is to cultivate an ecosystem where AI systems are engineered to be inherently ``safe by design'' (minimizing unintended harm) and concurrently ``secure by default'' 
(fortified against malicious exploitation). This dual focus is indispensable for fostering responsible innovation while safeguarding public trust.

\section{Conclusion}
\label{sec:conclusion}

The distinction between AI Safety and AI Security is critical for navigating AI development and deployment. AI Safety primarily focuses on preventing unintentional harm and ensuring AI systems operate as intended (``survival by design''). AI Security, conversely, defends AI systems against intentional, malicious threats (``defense by default''). 

While distinct, they are interdependent. Security lapses can lead to safety failures, as when attacks like prompt injection bypass an LLM's safety measures. Conversely, unaddressed safety vulnerabilities can become attack vectors for malicious misuse. Both concepts are ultimately human-centric.

By demystifying these terms and their boundaries, particularly clarifying how AI misuse and attacks relate to safety or security, we can foster more targeted research, develop effective governance, and ultimately build AI systems that are robustly safe and demonstrably secure. This clarity is paramount for realizing AI's full potential responsibly.

\section*{Acknowledgment}
We would like to thank Neil Gong, Xiaojun Jia, Byoungyoung Lee, Zhenkai Liang, Brendan Saltaformaggio, Carter Yagemann,  and Xu Yuan for helpful discussions and feedback on an early draft of the paper. 
This research was partially supported by NSF CAREER \#1942980, NSF CNS \#2112471, and the Schmidt Sciences Safety Science award.

\bibliography{references}

\begin{thebibliography}{51}
\providecommand{\natexlab}[1]{#1}
\providecommand{\url}[1]{\texttt{#1}}
\expandafter\ifx\csname urlstyle\endcsname\relax
  \providecommand{\doi}[1]{doi: #1}\else
  \providecommand{\doi}{doi: \begingroup \urlstyle{rm}\Url}\fi

\bibitem[Akuthota and Bhargava(2025)]{Apruzzese2023DeepLearningCybersecurityIDS}
Uday~Chandra Akuthota and Lava Bhargava.
\newblock The role of machine and deep learning in modern intrusion detection systems: A comprehensive review.
\newblock \emph{Computers and Electrical Engineering}, 124:\penalty0 110318, 2025.
\newblock ISSN 0045-7906.
\newblock \doi{https://doi.org/10.1016/j.compeleceng.2025.110318}.

\bibitem[Amodei et~al.(2016)Amodei, Olah, Steinhardt, Christiano, Schulman, and Man\'{e}]{Amodei2016ConcreteProblems}
Dario Amodei, Chris Olah, Jacob Steinhardt, Paul~F. Christiano, John Schulman, and Dan Man\'{e}.
\newblock Concrete problems in ai safety.
\newblock \emph{arXiv preprint arXiv:1606.06565}, 2016.

\bibitem[Anderson(2020)]{Anderson2020SecurityEngineering}
Ross~J. Anderson.
\newblock \emph{Security Engineering: A Guide to Building Dependable Distributed Systems}.
\newblock Wiley, 3rd edition, 2020.

\bibitem[Bai et~al.(2022)Bai, Kadavath, Kundu, Askell, Kernion, Jones, Chen, Goldie, Mirhoseini, McKinnon, et~al.]{Bai2022ConstitutionalAI}
Yuntao Bai, Saurav Kadavath, Sandipan Kundu, Amanda Askell, Jackson Kernion, Andy Jones, Anna Chen, Anna Goldie, Azalia Mirhoseini, Cameron McKinnon, et~al.
\newblock Constitutional ai: Harmlessness from ai feedback.
\newblock \emph{arXiv preprint arXiv:2212.08073}, 2022.

\bibitem[Barocas et~al.(2019)Barocas, Hardt, and Narayanan]{Barocas2019FairnessMLBook}
Solon Barocas, Moritz Hardt, and Arvind Narayanan.
\newblock \emph{Fairness and Machine Learning: Limitations and Opportunities}.
\newblock fairmlbook.org, 2019.
\newblock \url{http://www.fairmlbook.org}.

\bibitem[Bengio et~al.(2025)]{BengioEtAl2025InternationalAISafetyReport}
Yoshua Bengio et~al.
\newblock International {AI} safety report: The international scientific report on the safety of advanced {AI}.
\newblock Technical report, Produced with support from the UK Government, for the AI Safety Summit initiatives, January 2025.

\bibitem[Biggio et~al.(2012)Biggio, Nelson, and Laskov]{Biggio2012PoisoningSVM}
Battista Biggio, Blaine Nelson, and Pavel Laskov.
\newblock Poisoning attacks against support vector machines.
\newblock In \emph{Proceedings of the 29th International Conference on Machine Learning (ICML)}, 2012.

\bibitem[Bommasani et~al.(2021)Bommasani, Hudson, Adeli, Altman, Arora, von Arx, Bernstein, Bohg, Bosselut, Brunskill, et~al.]{Bommasani2021OpportunitiesRisksFoundationModels}
Rishi Bommasani, Drew~A Hudson, Ehsan Adeli, Russ Altman, Simran Arora, Sydney von Arx, Michael~S Bernstein, Jeannette Bohg, Antoine Bosselut, Emma Brunskill, et~al.
\newblock On the opportunities and risks of foundation models.
\newblock \emph{arXiv preprint arXiv:2108.07258}, 2021.

\bibitem[Bonnefon et~al.(2016)Bonnefon, Shariff, and Rahwan]{Bonnefon2016MoralMachine}
Jean-Fran{\c{c}}ois Bonnefon, Azim Shariff, and Iyad Rahwan.
\newblock The social dilemma of autonomous vehicles.
\newblock \emph{Science}, 352\penalty0 (6293):\penalty0 1573--1576, 2016.
\newblock \doi{10.1126/science.aaf2654}.

\bibitem[Bostrom(2016)]{Bostrom2014Superintelligence}
Nick Bostrom.
\newblock \emph{Superintelligence: Paths, Dangers, Strategies}.
\newblock Oxford University Press, 2016.

\bibitem[Brundage et~al.(2020)Brundage, Avin, Wang, Belfield, Krueger, Hadfield, Khlaaf, Yang, Toner, Fong, Maharaj, Koh, Hooker, Leung, Trask, Bluemke, O’Keefe, Koren, Ryffel, Rubinovitz, Besiroglu, Carugati, Clark, Eckersley, de~Haas, Johnson, Laurie, Ingerman, Krawczuk, Askell, Cammarota, Lohn, Krueger, Stix, Henderson, et~al.]{Brundage2020TowardTrustworthyAI}
Miles Brundage, Shahar Avin, Jasmine Wang, Haydn Belfield, Gretchen Krueger, Gillian Hadfield, Heidy Khlaaf, Jingying Yang, Helen Toner, Ruth Fong, Tegan Maharaj, Pang~Wei Koh, Sara Hooker, Jade Leung, Andrew Trask, Emma Bluemke, Cullen O’Keefe, Mark Koren, Th{\'e}o Ryffel, J.~B. Rubinovitz, Tamay Besiroglu, Federica Carugati, Jack Clark, Peter Eckersley, Sarah de~Haas, Maritza Johnson, Ben Laurie, Alex Ingerman, Igor Krawczuk, Amanda Askell, Rosario Cammarota, Andrew Lohn, David Krueger, Charlotte Stix, Peter Henderson, et~al.
\newblock Toward trustworthy ai development: Mechanisms for supporting verifiable claims.
\newblock \emph{arXiv preprint arXiv:2004.07213}, 2020.

\bibitem[Carlini and Wagner(2017)]{Carlini2017AdversarialAttacks}
Nicholas Carlini and David Wagner.
\newblock Towards evaluating the robustness of neural networks.
\newblock In \emph{Proceedings of the IEEE Symposium on Security and Privacy (S\&P)}, 2017.

\bibitem[Carlini et~al.(2021)Carlini, Tramer, Wallace, Jagielski, Herbert-Voss, Lee, Roberts, Brown, Song, Erlingsson, et~al.]{Carlini2021ExtractingTrainingData}
Nicholas Carlini, Florian Tramer, Eric Wallace, Matthew Jagielski, Ariel Herbert-Voss, Katherine Lee, Adam Roberts, Tom Brown, Dawn Song, Ulfar Erlingsson, et~al.
\newblock Extracting training data from large language models.
\newblock In \emph{30th USENIX security symposium}, pages 2633--2650, 2021.

\bibitem[Carlsmith(2022)]{Carlsmith2022ExistentialRiskPowerSeeking}
Joseph Carlsmith.
\newblock Is power-seeking ai an existential risk?
\newblock \emph{arXiv preprint arXiv:2206.13353}, 2022.

\bibitem[Cavoukian(2009)]{cavoukian2009privacy}
Ann Cavoukian.
\newblock Privacy by design: The 7 foundational principles.
\newblock \emph{Information and privacy commissioner of Ontario, Canada}, 2009.

\bibitem[Finlayson et~al.(2019)Finlayson, Bowers, Ito, Zittrain, Beam, and Kohane]{Finlayson2019AdversarialAttacksMedicalAI}
Samuel~G. Finlayson, John~D. Bowers, Joichi Ito, Jonathan~L. Zittrain, Andrew~L. Beam, and Isaac~S. Kohane.
\newblock Adversarial attacks on medical machine learning.
\newblock \emph{Science}, 363\penalty0 (6433):\penalty0 1287--1289, 2019.
\newblock \doi{10.1126/science.aaw4399}.

\bibitem[Floridi et~al.(2018)Floridi, Cowls, Beltrametti, Chatila, Chazerand, Dignum, Luetge, Madelin, Pagallo, Rossi, et~al.]{Floridi2021EthicsOfAI}
Luciano Floridi, Josh Cowls, Monica Beltrametti, Raja Chatila, Patrice Chazerand, Virginia Dignum, Christoph Luetge, Robert Madelin, Ugo Pagallo, Francesca Rossi, et~al.
\newblock Ai4people—an ethical framework for a good ai society: opportunities, risks, principles, and recommendations.
\newblock \emph{Minds and machines}, 28:\penalty0 689--707, 2018.

\bibitem[Fredrikson et~al.(2015)Fredrikson, Jha, and Ristenpart]{Fredrikson2015ModelInversion}
Matt Fredrikson, Somesh Jha, and Thomas Ristenpart.
\newblock Model inversion attacks that exploit confidence information and basic countermeasures.
\newblock In \emph{Proceedings of the 22nd ACM SIGSAC conference on computer and communications security}, pages 1322--1333, 2015.

\bibitem[Gabriel(2020)]{Gabriel2020AIValuesAlignment}
Iason Gabriel.
\newblock Artificial intelligence, values, and alignment.
\newblock \emph{Minds and Machines}, 30:\penalty0 411--437, 2020.

\bibitem[Ganguli et~al.(2022)Ganguli, Lovitt, Kernion, Askell, Bai, Kadavath, Mann, Perez, Schiefer, Ndousse, et~al.]{Ganguli2022RedTeamingLLMs}
Deep Ganguli, Liane Lovitt, Jackson Kernion, Amanda Askell, Yuntao Bai, Saurav Kadavath, Ben Mann, Ethan Perez, Nicholas Schiefer, Kamal Ndousse, et~al.
\newblock Red teaming language models to reduce harms: Methods, scaling behaviors, and lessons learned.
\newblock \emph{arXiv preprint arXiv:2209.07858}, 2022.

\bibitem[Goodfellow et~al.(2014)Goodfellow, Shlens, and Szegedy]{Goodfellow2014AdversarialExplaining}
Ian~J. Goodfellow, Jonathon Shlens, and Christian Szegedy.
\newblock Explaining and harnessing adversarial examples.
\newblock \emph{arXiv preprint arXiv:1412.6572}, 2014.

\bibitem[Greshake et~al.(2023)Greshake, Abdelnabi, Mishra, Endres, Holz, and Fritz]{Greshake2023MoreThanAMouthful}
Kai Greshake, Sahar Abdelnabi, Shailesh Mishra, Christoph Endres, Thorsten Holz, and Mario Fritz.
\newblock Not what you've signed up for: Compromising real-world llm-integrated applications with indirect prompt injection.
\newblock In \emph{Proceedings of the 16th ACM Workshop on Artificial Intelligence and Security}, 2023.

\bibitem[Gunning et~al.(2019)Gunning, Stefik, Choi, Miller, Stumpf, and Yang]{Gunning2019XAI}
David Gunning, Mark Stefik, Jaesik Choi, Timothy Miller, Simone Stumpf, and Guang{-}Zhong Yang.
\newblock {XAI}—explainable artificial intelligence.
\newblock \emph{Science Robotics}, 4\penalty0 (37), 2019.
\newblock \doi{10.1126/scirobotics.aay7120}.

\bibitem[Hendrycks et~al.(2019)Hendrycks, Zhao, Basart, Steinhardt, and Song]{Hendrycks2019NaturalAdversarial}
Dan Hendrycks, Kevin Zhao, Steven Basart, Jacob Steinhardt, and Dawn Song.
\newblock Natural adversarial examples.
\newblock \emph{arXiv preprint arXiv:1907.07174}, 2019.

\bibitem[Hendrycks et~al.(2022)Hendrycks, Carlini, Schulman, and Steinhardt]{hendrycks2022unsolvedproblemsmlsafety}
Dan Hendrycks, Nicholas Carlini, John Schulman, and Jacob Steinhardt.
\newblock Unsolved problems in ml safety, 2022.
\newblock URL \url{https://arxiv.org/abs/2109.13916}.

\bibitem[Hendrycks et~al.(2023)Hendrycks, Mazeika, and Woodside]{Hendrycks2023OverviewCatastrophicAIRisks}
Dan Hendrycks, Mantas Mazeika, and Thomas Woodside.
\newblock An overview of catastrophic ai risks.
\newblock \emph{arXiv preprint arXiv:2306.12001}, 2023.

\bibitem[Katz and Lindell(2007)]{katz2007introduction}
Jonathan Katz and Yehuda Lindell.
\newblock \emph{Introduction to modern cryptography: principles and protocols}.
\newblock Chapman and hall/CRC, 2007.

\bibitem[Koopman(2022)]{Koopman2017HowSafeIsSafeEnoughAV}
Philip Koopman.
\newblock \emph{How safe is safe enough?: Measuring and predicting Autonomous Vehicle Safety}.
\newblock 2022.

\bibitem[Koscher et~al.(2010)Koscher, Czeskis, Roesner, Patel, Kohno, Checkoway, McCoy, Kantor, Anderson, Shacham, and Savage]{Koscher2010AutonomousVehicleSecurity}
Karl Koscher, Alexei Czeskis, Franziska Roesner, Shwetak Patel, Tadayoshi Kohno, Stephen Checkoway, Damon McCoy, Brian Kantor, Danny Anderson, Hovav Shacham, and Stefan Savage.
\newblock Experimental security analysis of a modern automobile.
\newblock In \emph{Proceedings of the IEEE Symposium on Security and Privacy (S\&P)}, pages 447--462, 2010.

\bibitem[Leveson(2012)]{Leveson2011EngineeringSaferWorld}
Nancy~G. Leveson.
\newblock \emph{Engineering a Safer World: Systems Thinking Applied to Safety}.
\newblock The MIT Press, 2012.
\newblock \doi{10.7551/mitpress/8179.001.0001}.

\bibitem[Lundberg and Lee(2017)]{Lundberg2017SHAP}
Scott~M. Lundberg and Su-In Lee.
\newblock A unified approach to interpreting model predictions.
\newblock \emph{Advances in Neural Information Processing Systems}, 30, 2017.

\bibitem[M{\k{a}}dry et~al.(2018)M{\k{a}}dry, Makelov, Schmidt, Tsipras, and Vladu]{Madry2018DeepLearningModelsResistant}
Aleksander M{\k{a}}dry, Aleksandar Makelov, Ludwig Schmidt, Dimitris Tsipras, and Adrian Vladu.
\newblock Towards deep learning models resistant to adversarial attacks.
\newblock \emph{International Conference on Learning Representations (ICLR)}, 2018.

\bibitem[Marcus and Davis(2019)]{Marcus2019RebootingAI}
Gary Marcus and Ernest Davis.
\newblock \emph{Rebooting AI: Building Artificial Intelligence We Can Trust}.
\newblock Vintage Books, 2019.

\bibitem[Mehrabi et~al.(2021)Mehrabi, Morstatter, Saxena, Lerman, and Galstyan]{Mehrabi2021BiasFairnessSurvey}
Ninareh Mehrabi, Fred Morstatter, Nripsuta Saxena, Kristina Lerman, and Aram Galstyan.
\newblock A survey on bias and fairness in machine learning.
\newblock \emph{ACM Computing Surveys}, 54\penalty0 (6):\penalty0 1--35, 2021.

\bibitem[{National Institute of Standards and Technology}(2020)]{NIST_AC_SP800_53}
{National Institute of Standards and Technology}.
\newblock Security and privacy controls for information systems and organizations.
\newblock Technical Report Revision 5, U.S. Department of Commerce, September 2020.

\bibitem[{National Institute of Standards and Technology}(2023)]{NISTAIRMF}
{National Institute of Standards and Technology}.
\newblock Artificial intelligence risk management framework (ai rmf 1.0).
\newblock Technical report, U.S. Department of Commerce, 2023.

\bibitem[Orseau and Armstrong(2016)]{Orseau2016SafeInterruptibility}
Laurent Orseau and Stuart Armstrong.
\newblock Safely interruptible agents.
\newblock \emph{Proceedings of the Thirty-Second Conference on Uncertainty in Artificial Intelligence (UAI)}, 2016.

\bibitem[Ouyang et~al.(2022)Ouyang, Wu, Jiang, Almeida, Wainwright, Mishkin, Zhang, Agarwal, Slama, Ray, et~al.]{Ouyang2022RLHF}
Long Ouyang, Jeffrey Wu, Xu~Jiang, Diogo Almeida, Carroll Wainwright, Pamela Mishkin, Chong Zhang, Sandhini Agarwal, Katarina Slama, Alex Ray, et~al.
\newblock Training language models to follow instructions with human feedback.
\newblock \emph{Advances in neural information processing systems}, 35:\penalty0 27730--27744, 2022.

\bibitem[Qi et~al.(2024)Qi, Huang, Zeng, Debenedetti, Geiping, He, Huang, Madhushani, Sehwag, Shi, Wei, Xie, Chen, Chen, Ding, Jia, Ma, Narayanan, Su, Wang, Xiao, Li, Song, Henderson, and Mittal]{QiEtAl2024AIRiskFramework}
Xiangyu Qi, Yangsibo Huang, Yi~Zeng, Edoardo Debenedetti, Jonas Geiping, Luxi He, Kaixuan Huang, Udari Madhushani, Vikash Sehwag, Weijia Shi, Boyi Wei, Tinghao Xie, Danqi Chen, Pin{-}Yu Chen, Jeffrey Ding, Ruoxi Jia, Jiaqi Ma, Arvind Narayanan, Weijie~J. Su, Mengdi Wang, Chaowei Xiao, Bo~Li, Dawn Song, Peter Henderson, and Prateek Mittal.
\newblock {AI Risk Management Should Incorporate Both Safety and Security}.
\newblock \emph{arXiv}, May 2024.

\bibitem[Ribeiro et~al.(2016)Ribeiro, Singh, and Guestrin]{Ribeiro2016LIME}
Marco~Tulio Ribeiro, Sameer Singh, and Carlos Guestrin.
\newblock ``why should i trust you?'': Explaining the predictions of any classifier.
\newblock \emph{Proceedings of the 22nd ACM SIGKDD International Conference on Knowledge Discovery and Data Mining}, pages 1135--1144, 2016.
\newblock \doi{10.1145/2939672.2939778}.

\bibitem[Russell(2022)]{russell2019human}
Stuart Russell.
\newblock Artificial intelligence and the problem of control.
\newblock \emph{Perspectives on digital humanism}, 19:\penalty0 1--322, 2022.

\bibitem[Shafahi et~al.(2018)Shafahi, Huang, Najibi, Suciu, Studer, Dumitras, and Goldstein]{Shafahi2018PoisonFrogs}
Ali Shafahi, W.~Ronny Huang, Mahyar Najibi, Octavian Suciu, Christoph Studer, Tudor Dumitras, and Tom Goldstein.
\newblock Poison frogs! targeted clean-label poisoning attacks on neural networks.
\newblock \emph{Advances in Neural Information Processing Systems}, 31, 2018.

\bibitem[Shokri et~al.(2017)Shokri, Stronati, Song, and Shmatikov]{Shokri2017MembershipInference}
Reza Shokri, Marco Stronati, Congzheng Song, and Vitaly Shmatikov.
\newblock Membership inference attacks against machine learning models.
\newblock In \emph{Proceedings of the IEEE Symposium on Security and Privacy (S\&P)}, pages 3--18, 2017.

\bibitem[Souppaya et~al.(2022)Souppaya, Scarfone, and Dodson]{NIST_AISupplyChain_SP800_218}
Murugiah Souppaya, Karen Scarfone, and Donna Dodson.
\newblock Secure software development framework (ssdf) version 1.1.
\newblock \emph{NIST Special Publication}, 800\penalty0 (218):\penalty0 800--218, 2022.

\bibitem[Tanenbaum and Wetherall(2010)]{Tanenbaum2011ComputerNetworks}
Andrew~S. Tanenbaum and David~J. Wetherall.
\newblock \emph{Computer Networks}.
\newblock Prentice Hall, 5th edition, 2010.

\bibitem[Thacker()]{AISecuri39:online}
Joseph Thacker.
\newblock Ai security has serious terminology issues.
\newblock URL \url{https://josephthacker.com/ai/2023/10/16/ai-security-terminology-issues.html}.

\bibitem[Topol(2019)]{Topol2019DeepMedicine}
Eric Topol.
\newblock \emph{Deep medicine: how artificial intelligence can make healthcare human again}.
\newblock Hachette UK, 2019.

\bibitem[Tram{\`e}r et~al.(2016)Tram{\`e}r, Zhang, Juels, Reiter, and Ristenpart]{Tramer2016StealingModels}
Florian Tram{\`e}r, Fan Zhang, Ari Juels, Michael~K. Reiter, and Thomas Ristenpart.
\newblock Stealing machine learning models via prediction {APIs}.
\newblock In \emph{Proceedings of the 25th USENIX Security Symposium}, pages 601--618, 2016.

\bibitem[Uchida et~al.(2017)Uchida, Nagai, Sakazawa, and Satoh]{Uchida2017EmbeddingWatermarksNN}
Yusuke Uchida, Yuki Nagai, Shigeyuki Sakazawa, and Shin'ichi Satoh.
\newblock Embedding watermarks into deep neural networks.
\newblock In \emph{Proceedings of the 2017 ACM on international conference on multimedia retrieval}, pages 269--277, 2017.

\bibitem[Weidinger et~al.(2021)Weidinger, Mellor, Rauh, Griffin, Uesato, Huang, Cheng, Glaese, Balle, Kasirzadeh, et~al.]{Weidinger2021EthicalSocialRisksLLMs}
Laura Weidinger, John Mellor, Maribeth Rauh, Conor Griffin, Jonathan Uesato, Po-Sen Huang, Myra Cheng, Mia Glaese, Borja Balle, Atoosa Kasirzadeh, et~al.
\newblock Ethical and social risks of harm from language models.
\newblock \emph{arXiv preprint arXiv:2112.04359}, 2021.

\bibitem[Zou et~al.(2024)Zou, Phan, Wang, Duenas, Lin, Andriushchenko, Wang, Kolter, Fredrikson, and Hendrycks]{zou2024improvingalignmentrobustnesscircuit}
Andy Zou, Long Phan, Justin Wang, Derek Duenas, Maxwell Lin, Maksym Andriushchenko, Rowan Wang, Zico Kolter, Matt Fredrikson, and Dan Hendrycks.
\newblock Improving alignment and robustness with circuit breakers, 2024.
\newblock URL \url{https://arxiv.org/abs/2406.04313}.

\end{thebibliography}

\end{document}